# THE IMPACT OF WORLD WAR I ON RELATIVITY: A PROGRESS REPORT

By Virginia Trimble

University of California Irvine, Las Cumbres Observatory and

Queen Jadwiga Observatory, Rzepiennik, Poland

From an astronomical and relativistic point of view, the Great War began with the August, 1914 capture and imprisonment of the members of a German eclipse expedition that had gone to the Crimea to look, at the request of Einstein, for bending of starlight by the sun. And it ended in 1919 with the Eddington-inspired measurements of that light bending from Principe and Sobral and with the founding of the International Astronomical Union by scientists from "the countries at war with the Central Powers." In between came unprecedented, and in some ways unequaled, death and destruction. The scientists lost were mostly too young to have made an impact (Henry Moseley and Karl Schwarzschild are exceptions), but many of the best-known of the next generation had, if citizens of the belligerent countries, served on the battle lines, and most of the rest contributed in some other way. It may come as a surprise to find that both theoretical physics and observational astronomy of relevance to general relativity continued to take place and that there was a certain amount of communication of results, information, and even goods in both directions. The early post-war years saw something of a flowering of the subject, before the majority of physicists turned their attention to quantum mechanics and astronomers to stellar physics, though both had already been under consideration during the war. War-based bitterness between French and German scholars was surely part of the context in which Einstein and Henri Bergson faced off on 6 April 1922, in a debate on the nature of time, as part of an Einsteinian visit to Paris that had originally been planned for Fall, 1914.

**Introduction**

The decades on either side of 1900 saw a burgeoning of international activities in the science and life in general. Between 1900 and 1913, there were 428 international conferences in Paris, 168 in Brussels, 141 in London, and so forth[1]. The International Association of Chemical Societies got together in 1911 (in Paris, of course) and resolved to rationalize the names of organic compounds and to produce a fourth edition of Beilstein's catalogue[2]. The Association International des Académies was formalized in 1899 (in Weisbaden, despite the name). On the astronomical side, the AGK project had begun in 1868 and the Astrographic Catalogue ("Carte du Ciel") in 1887, under Admiral Mouchez at Paris[3].

The 1896 convention of the International Association for Geodesy was extended for a new decade in 1907 (their 1906 meeting in Budapest had introduced the rest of European earth-measures to the virtues of Eötvös Lorand's torsion balance). Kapteyn's Plan of Selected Areas received general agreement in 1906, following just after the establishment of the International Union for Co-operation in Solar Research, sparked by George Ellery Hale[4]. This last group was thinking, at the suggestion of Karl Schwarzschild, of expanding its remit to all of astrophysics at its 1910 meeting in Pasadena, but was still called the Solar Union for short at its last, August 1913 meeting in Bonn.

Since we shall not pass this way again, let us pause to note that Paul Jacobson, former professor at Heidelberg, who had established a commission in summer, 1914 to take on the chemical tasks, was not allowed to participate when the work actually got started under the International Union of Pure and Applied Chemistry[2]. And Henri Abraham, the founding general secretary of the International Union of Pure and Applied Physics, who was confirmed in office in 1934 and who had made (from Paris) significant contributions toward early radar in the First War, was deported and killed at Auschwitz during the Second.

**Those eclipse expeditions and how I came to the project**

Many years ago, I investigated whether scientific papers had grown monotonically longer more or less forever (the answer was a qualified yes)[5]. This was somehow in mind early in 2011 when I walked past the library shelves of *Nature* and noticed how skinny the volumes had become through the years 1915-19. A ha! Thought I: I wonder how else World War I shows up there. So I picked up the first August, 1914 issue and started to read, eventually examining every page 1914-19 and later back to 1908 and forward to 1923. Within a couple of issues, there was the report of a Berlin eclipse expedition to the Crimea under Erwin Freundlich having been captured and imprisoned. Paul Halpern[6] tells us more about the background, and issues of *Sirius*[*] and *Nature* more about the outcome.

World War II has been called the physicists' war (meaning radar, rockets, and fission bombs) and World War I the chemists' war (poison gases, of course, but also urgent need, in the face of various blockades, for nitrogen fixation, synthetic rubber and petroleum, dye stuffs, optical glass, and much else). Not surprisingly, in fact, no part of science, technology, or engineering came through either war unaffected. A handful of items that do not belong to us here include the first (accidentally) placebo-controlled test of vaccination (for typhoid fever), demonstration that round craters could come from oblique impacts, the first aircraft carrier, reluctant welcoming of women into industrial labs and production facilities, and a brief moment when Canada had the world's largest telescope. I have spoken about many such items at meetings on physics, chemistry, and astronomy, and hope some day to get it all on paper (or whatever anybody might be reading by then). This is the second installment. The first, "Who got Moseley's Prize?" (meaning the Nobel he had been nominated for in 1915 just before he was shot at Gallipoli, and the answer is Charles Barkla), is in an American Chemical Society book[7].

**The Prequels**

It will perhaps have occurred to you that the present author is not the first person to have noticed that we are currently passing through the centenaries of World War I (The Great War, before the second came along; July / August 1914 Sarajevo etc. to 18 June 1919, Versailles) and of general relativity (1913, the *Entwurf* theory to 1922

---

[*] *Sirius Zeitschrift fuer populaere Astronomia* is not very easy to find on the web (Sirius the star, Sirius XM radio and others dominate). I had never heard of it, and, when first thinking about this investigation, wondered whether the new *Naturwissenschaften* or the already old *Astronomische Nachrichten* might have some of the same kinds of information as *Nature*, but as seen from the other side. Neither journal was readily available, and I asked Hilmar Duerbeck (then retired from the University of Muenster, whom I knew slightly from his interest in "Who discovered Hubble's Law?") for advice and assistance. He responded quickly with xerocopies of a number of articles, 1915-1924, of *Sirius*, dealing with astronomers serving or killed on active duty, the trials and tribulations of the captured eclipse expedition, and the issue of rivalry between the new IAU and the old Astronomische Gesellschaft. Hilmar's sudden, unexpected, and very sad death on 5 January 2012 obviously put an end to our collaboration, and I still haven't done much of anything about information that might be found in *Naturwissenschaften* or *AN*, the latter of which had both obituaries and observatory reports. Hilmar also checked relevant issues of *Die Himmelswelt* and indicated that neither they nor *Jabresberichte* were very informative. A University of California Irvine colleague fluent in Russian, Meinhard Mayer, expressed some interest in what might have gone on under the Tsar, but died before we had done much except talk about the topic. I have not attempted since to enlist collaborators, feeling that it is somehow unlucky.

the first Friedman solution). The literature on each is enormous: something like 20,000 books on the war (of which I own about 0.3%) and 1700 books on Einstein[*]

Newton (1642-1727) on gravity and Clausewitz (1780-1831) on war is perhaps too far back to go for "origins," though the "causes" of the Great War can be traced back at least as far as the Treaty of Vienna, which more or less put an end to the Napoleonic Wars. The assassin of Archduke Franz Ferdinand who was the Trigger[8], does not seem to have been a very interesting person. The handful of tomes on the beginnings of the war that I have read (9-13) put forward so many causes and condemnations as to make it seem inevitable. This is the last I will say about WWI per se, except when it impinges on the creation of the general theory of relativity or related topics.

As for the origins of GR, Figure 1 is my own take and includes many of the outcomes as well. At a more technical, solemn, and serious level, something called *The Road to Relativity*[14] sounds (and is) particularly promising. The volumes by Kennefick[15] and Crelinstein[16] are Dr. Kormos-Buchwald's recommendation. The most succinct statement, made in print long ago, and reaffirmed this year by email, comes from Prof. Gerald Holton at Harvard, "only Einstein; only there; and only then," where "there" means Berlin, and "then" means 1914-16. Let's take them one at a time.

Biographers (read Pais[17] if you're having only one) and the man himself agree about Einstein's extreme ability to concentrate (sometimes to the exclusion of his family and friends, to put it very gently) and to keep after something until he saw it to be right, described as "the years of searching in the dark for a truth that one feels but cannot express, the intense desire and the alternation of confidence and misgiving until one breaks through to clarity and understanding, are known only to him who has experienced them himself"[18].

Berlin saw the arrival of Einstein on 29 March 1914 (from what is now the ETH in Zurich). He had been offered membership in the Royal Prussian Academy of Science, a quite decent salary for the time and place, and no teaching responsibilities. Wife Mileva and sons Hans and Eduard came with him briefly, but soon returned to Zurich; among the other attractions of Berlin was the presence of Albert's cousin Elsa, whom he later married. Despite the "no teaching" arrangement, by 1917 he appears in a *Physikalische Zeitschrift* list of lectures to be given that year at Berlin, among those at more than a dozen German-speaking universities (including Vienna and Zurich), with the topic "relativity." In fact, he also lectured on quantum theory and statistical mechanics at various times. Emmy Noether also appears in the list, lecturing at Göttingen. (19, 20 pp. 561, 735 eg).

Einstein left behind friend and collaborator on the *Entwurf* theory (of which more shortly) Marcel Grossmann (1878-1936), but came to a large collection of physicists in Germany. Because Switzerland (and Holland) remained neutral during that war, he was able to continue to communicate with, collaborate with, and even visit physicists and mathematicians there.

---

[*] Diana Kormos Buchwald, director of the Einstein Papers Project in Pasadena reports that there are something like 1,700 books on Einstein, of which about 200 are biographies, the first in 1921; and she should know!" And let me pause right here to acknowledge that she most generously sent me copies of the three volumes of "Einstein papers"[20] that contain all the personal letters to and from him that have been found for the years 1914-1918. These will appear repeatedly below.

The men whom Gutfreund & Renn regard as "influential" for Einstein and who belong to the "before and during GR" period include the following, (award yourself one Brownie Point for each name you can honestly say you already knew at least a little bit about – my score was 15 before beginning this project):

In Zurich: Paul Bernays (1888-1977), Michele Besso (1873-1955), Marcel Grossmann (1878-1936), Hermann Weyl (1885-1955, later Göttingen and IAS Princeton)

In Leiden: Paul Ehrenfest (1880-1933), Hendrik Lorentz (1853-1928), Willem de Sitter (1872-1934)

In Italy: Tulio Levi-Civita (1873-1943), Padua, later Rome), Max Abraham (1875-1922, Milan, returned to Germany at declaration of war, war work, later Stuttgart)

In Austria-Hungary: Lorand Eötvös (1848-1919, Budapest), Friedrich Kottler (1886-1965, Vienna), Ernst Mach (1838-1916, Vienna to 1901, member of parliament 1901-13, retired at home near Vienna 1913-16)

In Russian territory: Gunnar Nordström (1881-1925, Helsinki, some time in Copenhagen)

In Germany: Max Born (1882-1963, Göttingen to radio operator to sound ranging), David Hilbert (1862-1963, Göttingen) Gustav Mie (1869-1957, Griefswald to Halle-Wittenberg in 1917), Hermann Minkowski (1864-1909, Göttingen), Walther Nerst (1864-1941, Berlin), Arnold Sommerfeld (1868-1951, Munich), Max von Laue (1879-1960, Frankfurt, later Berlin)

You might want to add any number of footnotes, for instance that you associate Mie and Nernst primarily with other parts of physics (more then than now, lots of creative scientists worked on many topics); that some of these folks won their own Nobel Prizes, and some didn't; that as early as 1913 Einstein had asked George Ellery Hale about measuring light deflection in the sun's gravitational field; that the elder Bragg was also working on sound ranging on the Allied side while the younger one was on the front lines…Out, bad dog, this belongs to a different WWI story!

Thirdly comes the time frame. I find it much harder to conceive of this as favorable, let alone uniquely favorable. Early in the war, Einstein declined to sign a statement by 93 other German savants denying responsibility for starting the war, and indeed was briefly involved in discussions of trying to provide a contrary document (there were not many takers)[21]. Later, inflation began to erode that salary, and a great deal of time went into trying to arrange travel back to Zurich to see colleagues and his sons and to Leiden. In due course, more and more effort had to go into organizing finances and, eventually, food. He was in fact quite ill through portions of 1917-18, though that was after the crucial time period for conceiving, working out, and publishing the basics of general relativity. To realize how extreme things became, even for the "privileged classes," just read the letters in ref. 20. Well, you can skip the equations for now, though we will need them later. On the other hand, there were surely many fewer requests for more distant travel, speeches, conference participation, and so forth than either before or after the Great War.

Prewar, he appears in at least a couple of Solvay conference photographs, and in 1922 he made the visit to Paris postponed from fall 1914.* Much-photographed trips to Japan (where he learned of his Nobel Prize) and the United States followed the war, then Tel Aviv, of which he was made an honorary citizen, and where he spoke briefly in Hebrew then continued in French, en route home from Japan. So perhaps indeed, only "there" and only "then."

**The Entwurf Theory**

*Entwurf* is, I think, a lovely word, blending the Entmoot of *Lord of the Rings* with a greeting from Einstein's beloved canine, Chico[22]. Co-author Marcel Grossmann was the only person thanked in Einstein's special relativity paper, and "Entwurf" is one of two significant gravitation papers by the pair[25]. The full title is "Entwurf verallgemeinerten Relativistätstheorie und einer Theorie der Gravitation." In fact the word means sketch, outline, or draft, and is a compound of the German "ent" indicating entry into a new state and "wurf" meaning throw, cast, projection, or direction[26].

What is in it? We'll peek in just a moment, but this was not Einstein's first attempt at generalizing his special theory to include accelerated observers. The story is told very concisely in Misner, Thorne, and Wheeler[27], section 17.7, at considerable length in the four volumes of *The Genesis of General Relativity*[28], with pedagogical intent by Dwight Neuenschwander[29], and at "Goldilocks" length and level by Gutfreund and Renn[14].

The 1908 paper[30] "On the relativity principle and the conclusions one draws from it" starts by supposing the equivalence of acceleration and gravitation. This is customarily illustrated with a small person in an elevator and accompanied by some mention of the experiments of Eötvös Lorand (or Roland Eötvös if you prefer). Eötvös had indeed introduced a number of European scientists to his work when an international geodesy society met in Budapest in 1906. In 1911 comes "Über ein Einfluss der Schwerkraft auf die Ausbreitung des Lichtes"[31].* That "influence on the propagation of light" is, of course, what we now call bending of light and counts as one of three (or so) classic tests of GR. This early prediction, and the "Entwurf" one, were, however, for angles half of what is now calculated from the 1915-16 general theory. The result, 0.87" or thereabouts, at the limb of the sun is numerically equal to (though I think intellectually distinct from) the Newtonian number found by Cavendish and Soldner (and a slightly different path from here leads us into gravitational lensing).

The year 1912 saw two papers and a note in proof on static gravitational fields and the question "Is there a Gravitational Effect which is analogous to Electrodynamic Induction?" The last of these enables us to answer the question "Was Einstein a Machian?" with a good, firm, "sometimes." Meanwhile, others were treading on his heels (and possibly also toes) – Max Abraham[33] with his own theory of gravity, incorporating Minkowski's formalism (in a way that Einstein criticized and Abraham modified), and Gunnar Nordström[34] claiming consistency with special relativity (briefly regarded with favor by Einstein, but later[35] disputed, in one of his relativity rare joint papers with someone not now greatly remembered. The co-author, Adriaan D. Fokker, was Dutch and a cousin of Anthony Fokker, caught in Germany and put to work designing aircraft with a gubbins to synchronize propeller and machine gun so that the pilot did not shoot off his own blades.

---
* The original invitation had come from Longevin, and the spring, 1914 visit was apparently uneventful. But 6 April 1922 degenerated into a nasty squabble with French philosopher Henri Bergson. Some details appear in a later section, but if you can't wait, check out refs. 23 and 24. To first order, Canales is pro-Bergson, Topper pro-Einstein. Echoes of the war affected their interactions and their colleagues' reaction to it.
* If you switch back and forth between the English and German titles often enough, you eventually stop noticing which is which. This probably says something good about Einstein's writing style, because I did not have the same experience when tackling Theodor Storm's Immensee some years ago.

What is in it? The *Entwurf* theory is an attempt to formulate equations that show how the gravitational field is related to the distribution of mass-energy that creates that field. It makes use of the Riemann tensor (a contribution of Grossmann), but not of the Christoffel symbols, but (1) it did not satisfy a correspondence principle in the sense of reducing to Newtonian gravity under suitable conditions (we are typically more used to thinking of "correspondence principles" as pertaining to a classical limit of quantum mechanics). (2) It led to an advance of the perihelion of Mercury equal to one-half the observed number (so we should not be surprised that it also predicted bending of light half that of the later theory), (3) it did not permit treating a rotating reference frame as equivalent to a system at rest (a failure of Mach's principle in one of its many meanings), (4) it was not fully covariant, which Einstein still thought would be a "good thing" because of its association with energy-momentum conservation, though he had apparently shown that generally covariant theories would violate causality.

In case you hadn't thought of them together lately, Einstein was one of the people who recognized very early the outstanding quality of Emmy Noether's mathematical ability and the significance of her best-known theorem about symmetries and conservation laws.

**Back to the Calculating Board: October 1914 to December 1915**

The dates are from the submission of the *Entwurf* paper to that of the last of the four "now I've got it!" GR papers. What was Einstein doing; who else was involved; and what difference did the war make? We make contact with the outbreak of the war via a footnote to letter 34 in (20) to Ehrenfest (19 August 1914) indicating that the value of the items Fruendlich left behind was 20,000 marks for a Zeiss telescope and 2086 marks for other items. Albert then turns his attention to assuring his son Hans Albert (Lieber Albert) that he is packing up everything to send to the three of them in Zurich (10 September) and explaining to L(ieber) M(ileva) that he can send no more money, because he has no more until he is paid again. His regular salary was 900 marks per annum (paid quarterly in arrears) and a special personal salary of 12,000 marks, at least the first year. He later speaks of himself as living the simplest, almost meager existence, having to pay for his mother's cancer operation, and so forth.

Issues of money, and increasingly, food appear scattered through the science[20]. Some of it sounds a bit petty, but I think Mileva must have been terrified at having sole responsibility for two sons, one of whom, it was already clear, was going to be a "special child."

Both international and academic politics take up a good deal of his time during this critical window. Einstein is asked for advice on how a Swiss university should fill an empty chair (since German candidates were not appropriate during hostilities and no Swiss candidate obviously stands out). We might think "France? Britain" but they (and he) did not, and he tries repeatedly to get Erwin Freundlich reappointed from a route task in stellar astronomy to a position where he can contribute to observational tests of relativity, particularly gravitational redshift in binary stars and light bending by Jupiter. We might think this latter impossible, and Einstein thought it nearly unnecessary, with GR adequately supported by the perihelion of Mercury calculation and gravitational redshift of the sun.

On the international front, he regards the war, from the beginning as ample folly (and other harsher words), for which His (God's) nonexistence alone can excuse Him (Doc 44)[20]. Einstein hopes initially that scientists from the various combatant countries might at least retain some support for internationalism (while reminding those who want to coordinate such writing or activity that he is a Swiss national not a German)[*] but soon realizes his colleagues are no more open-minded on this than the general population. Invoking a hero from the past, an organization called the Berliner Goethesbund asked for a supporting statement from Einstein. He drafted a couple but was apparently unable to satisfy them. Personally, he was very much incommoded by the difficulty of crossing even the neutral Dutch and Swiss borders to meet with colleagues and visit his sons in Zurich. The difficulties and Einstein's disconnect from the majority of his war-supporting colleagues of course intensify through to the end in 1918, as do the problems of food, money, and illness. Yes, he has the flu in 1918 as do family and friends.

Whatever the original arrangement may have been Einstein in fact gave lecture courses on relativity in the fall of 1914-15 and summer 1915 semesters. He adjusted the timing of the latter in order to visit his sons in Switzerland, the elder of whom, however, did not desire the visit. So it did not happen.

Turning to science, I found it a surprise that relativity was not the only issue Einstein was thinking about during this critical "Entwurf-to-triumph" period. Topics that turn up in the correspondence include the design and operation of a planimeter, the spectrum of helium, the Bohr atom, the atmospheric absorption of light, and the possible existence of a magneton (quantum of magnetic field). The most important, however, were (1) entropy and the issue in thermodynamics of whether one can ever reach 0 K in a finite set of operations (letters mostly to and from Michael Polyani, and later with Planck and Ehrenfest, concerning some formulae due to Tetrode, who was a person much though he sounds like a tensor), and (2) an on-going series of experiments with Wander de Haas (1878-1960).[*]

What were they measuring, or attempting to measure, and why? Ampere's molecular currents. At least 19 of the 1914-18 letters touch on the subject; many also address Einstein's efforts to help the de Haas family move from Berlin to the Netherlands, and getting their furniture sent after them. Einstein remarks that to help someone move during war time is a test of true friendship! As for the physics, Ampere is associated with the force between a pair of current carrying wires. The experiment put a thin iron rod in a magnetic field maintained by coils parallel to it. The spinning electrons act like little magnets and line up with the field. Switch the field back and forth to excite oscillations in the torque on the rod, and the measured period and amplitude tell you the charge to mass ratio of the electron (Millikan did this differently!). Not surprisingly, the main problem is correcting for all the other forces and effects around, and Einstein in effect signs the work over to de Haas in May 1915 (letter 82), a couple of papers being published in due course[36]. In spite of the difficulties, they did firmly establish that the current carriers have negative charge. You knew that? Yes, but the equations for electricity were originally written down on the assumption that the carriers had plus charges, as defined by Benjamin Franklin long ago. In the "relations were becoming both strained and numerous" department, de Haas was also the son-in-law of Hendrik Larentz (1853-1928), who took some interest in the analysis of the experimental data (letter 79) and was also in the same time frame in correspondence with Einstein about tensors in/and relativity.

---

[*] He later resumed German citizenship.

[*] "Have I head of this guy?" De Haas is possibly slightly more familiar as part of the de Haas-van Alphen effect, which dates from 1930, was predicted by Lev Landau also in 1930 and fully explained by Lars Onsager only in 1952. The words from Wiki say that it is the oscillation of the magnetic moment of a pure metal crystal when an applied magnetic field increases (at level of 2-3 tesla). The resistivity, specific heat, and sound attenuation also oscillate. Pieter van Alphen was De Haas's student. The Einstein-De Haas effect (that oscillation of torque in the iron rod) is also somewhat quantum mechanical, in the sense that to calculate the right response to a given applied field, you need a Lande g factor near 2 and Quantum ElectroDynamics. Now the only puzzle is the name. Why Ampere? Why molecular (when apparently individual atoms and their electrons are involved)? And why current? V. Ya. Frenkel wrote on the topic in 1979 in case you remember him from other things I have written.

And now, the essential work that turned *Entwurf* into the four papers presented to the Berlin Academy on November 4, 11, 18 and 25, 1915. These are all Thursdays, as were the dates generally given for his submission of *Entwurf* (29 November 1914), Schwarzschild's solution (24 Feb 1916), of which Schwarzschild had informed Einstein by letter, and the first paper mentioning gravitational waves (22 June 1916). Apparently it is just that The Berlin Academy met on Thursdays, and reading a paper to them counted as submission for the proceedings. The Paris Academy also continued its regular weekly meetings through the War, and you can read summaries of what went on in *Nature*. For the Berlin Academy, you must go to the German literature. And by 9 December (letter 161) Einstein can assure Sommerfeld that the four November papers are the most valuable finding he has made in his life (or would ever make, we would now say). He also describes them as the final stage in the battle over the field equations, being fought out before your eyes!

**Facing up to General Relativity**

This brings us to GR and what Einstein did between the *Entwurf* (of which he had almost immediately doubts) and the triumph.

First he needed a metric. That's the entity that is the same for all observers allowed to exist in the theory. For flat Cartesian surfaces, it is

$$ds^2 = dx^2 + dy^2 \quad \text{(or the square root there of)},$$

the distance between two points on the plane. Adding time, as in special relativity, makes it the distance between two events.

Hermann Minkowski (1864-1909) had shown in 1907 that the invariant interval in special relativity could be expressed geometrically as

$$ds^2 = c^2 dt^2 - (dx^2 + dy^2 + dz^2)$$

the Minkowski metric. Notice that $ds^2$ means $(ds)^2$ not a second derivative, and that, with this sign convention, positive values of $ds^2$ describe pairs of events that you can get between (travelling and $v = c$ or less) in time to experience both. $ds^2 = 0$ is a photon or other massless particle that travels at $v = c$.

The full three space + one time dimension equivalent is $ds^2 = g_{\mu\nu} dx^\mu dx^\nu$, where $\mu$ and $\nu$ run over 0 (time) 1,2,3 (space) and we meet the Einstein summation convention, that an index that appears both "upstairs" and "downstairs" is to be summed over. Einstein wrote his metrics and such that way, and his mathematical notation is much easier to read than his handwriting (see ref 14, which reproduces one of his manuscripts). The metric was well in place before *Entwurf*.

His goal was to express the geometry of space time (the $g_{\mu\nu}$ tensor) as a function of the distribution of mass-energy in the space-time to be modeled. Clearly this also has to be a tensor, generally called $T_{\mu\nu}$. As before, the index names can be any letters you want, and you can choose to make either time-like or space-like intervals between the events the ones with positive $ds^2$ (textbooks differ on the choice). $T_{\mu\nu}$ will have components for density of matter and energy, for fluid flows, for various stresses, and so forth. It is symmetric, so that there are at most 10 independent pieces, not 16. Electromagnetic energy and momentum count too, and yield coupled Einstein-Maxwell equations (but this is not a unified theory).

The *Entwurf* theory had the stress-energy tensor. "Moving right along" he then needed an equation of motion describing the motion of particles in a given gravitational field, and a field equation, describing the gravitational field generated by its matter and energy sources. The former was in place by 1912. The latter presented more serious problems.

Einstein was sure that the proper field equations would have several properties: (1) reducing to the Newtonian solution when fields were weak, (2) treating gravitation and acceleration as interchangeable (the equivalence principle), (3) conserving energy and momentum, and (4) full covariance.

*Entwurf* was OK on (1) and (2). But the form he had chosen for the tensor (a function of $g_{\mu\nu}$) representing the geometry of space time resulted in field equations that, if forced to be covariant, did not conserve energy and momentum (or conversely).

We absorbed conservation with whatever our earliest caregivers fed us. But what is covariance? Einstein wrote:

> "The general laws of nature are to be expressed by equations which hold good for all systems of coordinates, that is, are covariant with respect to any (coordinate) substitution whatever (generally covariant)"

At this point, I am plagiarizing freely from ref 37[*], on the grounds that, whatever authorial copyrights still exist are now mine as sole heir and executrix.

Let's try once more. You have the physical system you want to study (the $T_{\mu\nu}$) and what you think are the right field equations, $T_{\mu\nu}$ = some function of $g_{\mu\nu}$. You solve the equations in some coordinate system and predict how a particle should move in that physical set up. You transform to another coordinate system and oops, you no longer have a solution of the field equations; your theory is not fully covariant. Notice that Einstein insisted upon retaining conservation when, in the *Entwurf* theory, it conflicted with full covariance. Something must have gone wrong with his $G_{\mu\nu}$, the function of the metric $g_{\mu\nu}$ to be set equal to $T_{\mu\nu}$.

At *Entwurf* time, he knew (thanks to the mathematical expertise of Marcel Grossmann) about the Riemann and Ricci tensors (which can describe the curvature of space) and the work of Levi-Civita, and made use of them. Through the critical 10 months before the November papers, there was extensive correspondence between Einstein and Levi-Civita, Hendrik Lorentz, and Hilbert (de Sitter comes a bit later) about the proper ways to write and use

---

[*] This source makes time-like intervals positive, though I don't think I knew that when I married him.

tensors in theories of gravity. The "tone of ink" fluctuates. He is first deeply impressed by and attracted to Hilbert; then for a time regards him as an idea-thief, and then again later as a friend. There are to Lorentz 14 letters, Hilbert 16, Levi-Civita 12 (ending with a sort of "this correspondence is now closed"). It is clear from the letters and postcards (which apparently crossed the Swiss and Dutch borders more readily than sealed letters) which survive that others, perhaps many, do not, having been discarded by their recipients, the heirs, or non-relativists, interested only in cutting off the autograph.[*]

But it is a 30 September 1915 note to Erwin Fruendlich (Doc 123) concerning rotating systems that reveals the contradiction clearly for us bears of very little brain (which Freundlich was not!!). Using a $T_{\mu\nu}$ for a slowly rotating coordinate system and the last (time coordinate) of the *Entwurf* equations, he finds

$g_{44} = 1 - (3/4) \omega^2 (x^2 + y^2)$

while the direct transformation from the Galilean case yields

$g_{44} = 1 - \omega^2 (x^2 + y^2)$. He therefore supposes that the calculation for the perihelion of Mercury (18"/century) is suffering from the same fault, but is not sure whether it is the field equations themselves or his application that caused the discrepancy. Implicitly the *Entwurf* prediction for light bending will have some similar error. As it happens, weak gravitational redshifting is not affected.

The secret word, as Einstein wrote to Sommerfeld on 28 November 1915 is "Christoffel'sche Tensor", translated as Christoffel's symbols (doc 153), and indeed they are not tensors in terms of how they transform between coordinating systems.

The notation of the letter is not quite modern, but returning to Ref. 37, we find the field equations given as

$(8 \pi G/c^4) T_{\mu\nu} = R_{\mu\nu} \frac{1}{2} g_{\mu\nu} R$ where

$g_{\mu\nu}$ is the original metric tensor you thought of;

$R = g^{\mu\nu} R_{\mu\nu}$   (there are rules for raising and lowering indices)

$R_{\mu\nu} = R^{\alpha}{}_{\mu\alpha\nu} = \frac{\partial \Gamma^{\alpha}{}_{\mu\nu}}{\partial x^{\alpha}} - \frac{\partial \Gamma^{\alpha}{}_{\mu\alpha}}{\partial x^{\nu}} + \Gamma^{\alpha}{}_{\mu\nu}\Gamma^{\beta}{}_{\alpha\beta} - \Gamma^{\alpha}{}_{\mu\beta}\Gamma^{\beta}{}_{\nu\alpha}$ (called the Ricci tensor)

$R^{\mu}{}_{\alpha\beta\gamma} = \frac{\partial \Gamma^{\mu}{}_{\alpha\gamma}}{\partial x^{\beta}} - \frac{\partial \Gamma^{\mu}{}_{\alpha\beta}}{\partial x^{\gamma}} + \Gamma^{\mu}{}_{\sigma\beta}\Gamma^{\sigma}{}_{\alpha\gamma} - \Gamma^{\mu}{}_{\sigma\gamma}\Gamma^{\sigma}{}_{\alpha\beta}$

(the Riemann, Riemann-Christoffel or curvature tensor; it is really big when space is greatly distorted by mass, and blows up at singularities)

and

$\Gamma^{\gamma}{}_{\mu\alpha} = \frac{1}{2} g^{\gamma\nu} \left[ \frac{\partial g_{\nu\mu}}{\partial x^{\alpha}} + \frac{\partial g_{\nu\alpha}}{\partial x^{\mu}} - \frac{\partial g_{\mu\alpha}}{\partial x^{\nu}} \right]$

Einstein wrote the Christoffel symbols as $\{ \frac{\gamma}{\mu\alpha} \}$

And that's all there is to general relativity, except the detail that the physical system you want to study gives you $T_{\mu\nu}$ and you have to run around guessing metrics $g_{\mu\nu}$ until you find one which, subjected to this massive mathematical manipulation, gives you back the $T_{\mu\nu}$ you want.

---

[*] Signatures/autographs of Albert Einstein, unattached to letters with scientific content, retail for prices like $300-$500, more than George Bernard Shaw (who wrote a great deal) but much less than Richard Wagner (who wrote mostly music).

All of us who have taught or taken a required course in GR (and learned that it is more blessed to give) will agree that the underlying idea of a relationship between the distribution of mass-energy and the geometry of space time is a straightforward one, but the arithmetic is a bit grim. Einstein himself, wrote that some place, and also expressed to Karl Schwarzschild (20) (doc 176, 181) surprise that the exact solution for a point mass could be formulated so simply, mentioning that the computational problems are inordinately large, and that he proposed to present Schwarzschild's work to the academy next Thursday (Jan 13 1916). The next month Einstein informs Schwarzschild that there are no gravitational waves analogous to electromagnetic waves (that is dipole, and of course correct – the reality of quadrupole waves belongs to another story) (20, doc 194).

This concludes Part I. Part II, currently at the Klein-bottle stage, will look in (perhaps excessive) detail at many of the other scientists and mathematicians who interacted with Albert Einstein, General Relativity, and, inevitably, the Great War in the period 1913-19. Part III, currently being assembled, deals with the aftermath of both GR and WWI, primarily in the time frame 1919-1926.

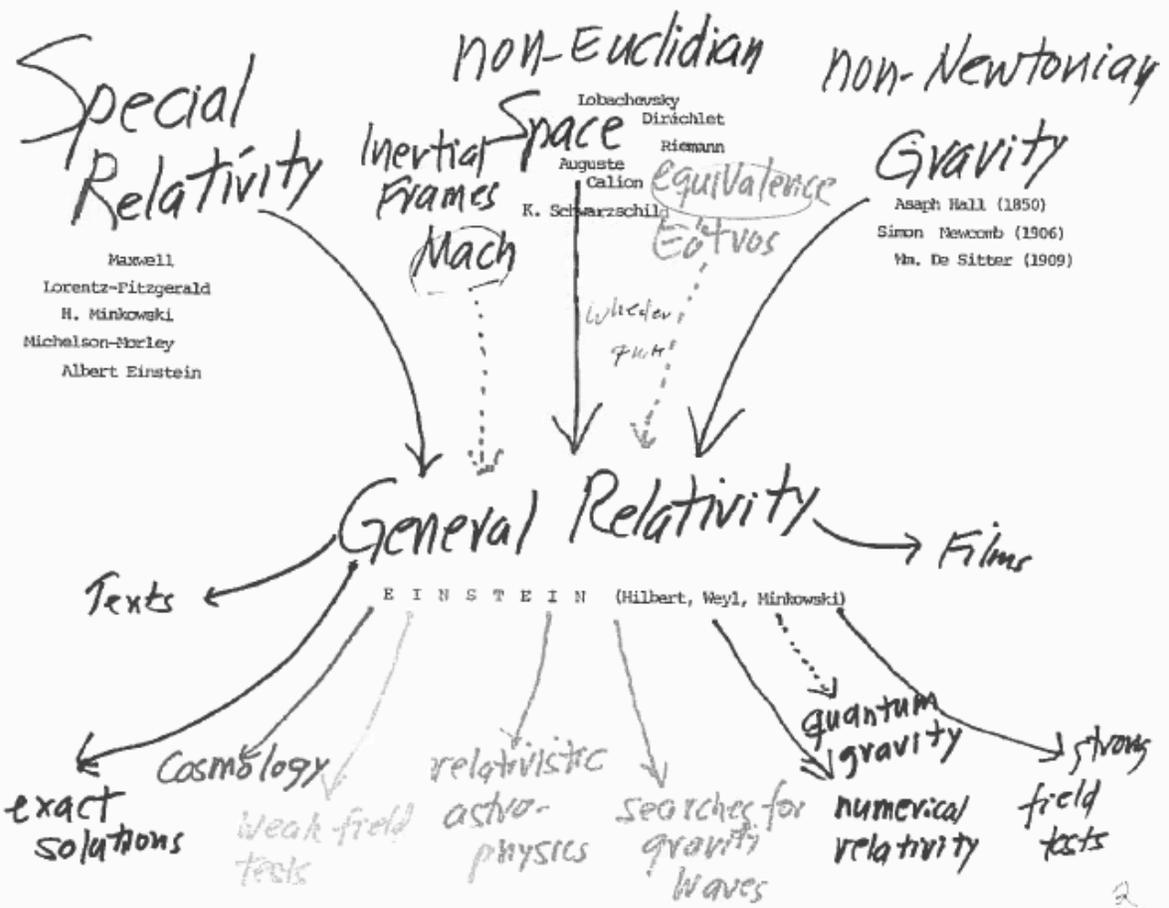

Figure 1: My take on where GR came from and what led out of it